# Thermal Expansion and Diffusion Coefficients of Carbon Nanotube-Polymer Composites

Chenyu Wei *

*NASA Ames Research Center, MS 229-1, Moffett Field, California 94035*

*Department of Mechanical Engineering, Stanford University, California 94305*

Deepak Srivastava

*NASA Ames Research Center, MS T27A-1, Moffett Field, California 94035*

Kyeongjae Cho

*Department of Mechanical Engineering, Stanford University, California 94305*

*Corresponding author: cwei@stanaford.edu, NASA Ames Research Center, MS 229-1, Moffett Field, California 94035

**ABSTRACT:** Classical molecular dynamics (MD) simulations employing Brenner potential for intra-nanotube interactions and Van der Waals forces for polymer-nanotube interfaces are used to invetigate the thermal expansion and diffusion characteristics of carbon nanotube-polyethylene composites. Additions of carbon nanotubes to polymer matrix are found to increase the glass transition temperature $T_g$, and thermal expansion and diffusion coefficients in the composite above $T_g$. These findings could have implications in CNT composite processing, coating and painting applications.

Carbon nanotubes (CNTs) are nano-scaled materials that are found to have unusual mechanical and electronic properties. [1] The strong in-plane graphitic C-C bonds make them exceptionally strong and stiff against axial strains and very flexible against non-axial strains. Additionally, CNTs can be metallic or semiconducting, determined by their atomic configurations such as chirality and diameter. Many applications of CNTs, such as in nano-scale molecular electronics, sensing and actuating devices or as reinforcing additive fibers in functional composite materials, have been proposed. Several recent experiments on the preparation and mechanical characterization of CNT-polymer composites have also appeared. [2] [3] [4] These measurements suggest modest enhancements in strength characteristics of CNT-embedded matrixes as compared to bare polymer matrixes. [4] Preliminary experiments and simulation studies on the thermal properties of carbon nanotubes show very high thermal conductivity. [5] It is expected, therefore, that nanotube reinforcements in polymeric materials may also significantly change the thermal and thermo-mechanical properties of the composites.

Using classical molecular dynamics (MD) simulations (DLPOLY MD program [6]), we investigate the thermo-structural behavior of CNT reinforced polyethylene (PE) composites. The structural and diffusion behavior of the composites has been studied in the temperature range below and above glass transition temperature $T_g$. We report significant enhancements of thermal expansion and diffusion coefficients of the composite materials above $T_g$.

The interactions of the carbon atoms within nanotubes are described by Tersoff-Brenner many-body potential, [7] [8] which is fitted to describe carbon and hydrocarbon systems. This potential has been used to study the elastic properties of carbon nanotubes and given results comparable to that from Tight Binding method or Density Functional Theory. The nanotube-polymer interface is described via non-bonding Van der Waals interactions (VDW) of truncated Lennard-Jones (LJ) 6-12 type with $\varepsilon = 0.461$ kJ/mol and $\sigma = 3.65$ Å. The united-atom model is used to describe individual polymer chain, in which each CH2 or CH3 group is considered as a single interaction unit. This model is commonly used to avoid the high frequency vibrations of C-H bonds. We adopt the form of intra-polymer potentials used by Clarke et al. [9] The interactions within a polymer chain have two components: valence angle potential and torsion potential. The valence angle potential is described as:

$$\Phi(\theta) = 0.5 k_\theta (\cos\theta - \cos\theta_0)^2,$$

where $k_\theta = 520 \text{J} \cdot \text{mol}^{-1}$, $\theta_0 = 112.813°$. The torsion potential is described as:

$$\Phi(\alpha)/\text{J}\cdot\text{mol}^{-1} = C_0 + C_1 \cos\alpha + C_2 \cos^2\alpha + C_3 \cos^3\alpha,$$

where $C_0 = 8832$, $C_1 = 18087$, $C_2 = 4880$, and $C_3 = -31800$. The C-C bonds in a polymer chain can be kept rigid to have large time steps. However, we allow full dynamics of these bonds, to be consistent with the C-C bonds within CNTs. A harmonic potential $0.5 k_b (l-l_0)^2$ is used for the C-C bonds within a chain, where $k_b = 346 \text{KJ} \cdot \text{mol}^{-1}/\text{Å}^2$ and $l_0 = 1.53$ Å. [10] Truncated 6-12 LJ potentials are applied to pairs of units parted by more than three units with $\varepsilon = 0.498 \text{KJ} \cdot \text{mol}^{-1}$ and $\sigma = 3.95$ Å. The time-step of 0.5fs is used in our MD simulations.

The conformation of a long-chained polymer molecule is important in describing the dynamics and properties of bulk polymers. [11] Studies have shown that MD is very slow to relax a polymer chain to its equilibrium conformation. This difficulty can be avoided by using Monte Carlo (MC) simulations. In our study MC simulations are run on each chain at 300K for up to two million steps, such that the end-to-end distances of polymer chains show the square-root dependence on chain lengths, in agreement with the prediction of Flory's theory. [12] MC relaxed chains are then used in MD equilibrations of bulk configurations with periodic boundary condition. The initial overlapping of chains is removed by gradually turning on the VDW interactions.

We first investigate small size samples. For bulk PE, samples are consisted of 80 short chains, with repeating units $N_p = 10$, in each unit cell (total 800 atoms). For CNT composite samples, each unit cell has a 20 Å long capped (10,0) CNT (total 204 atoms) embedded in the polymer matrix, and the volume ratio of CNTs is about 8%. Periodic boundary condition is used and an illustration of the simulation unit cell for the composite is shown in Figure 1. The initial unit cell size is 28 Å × 28Å × 30 Å for both polymer and composite system. The Evans NVT ensemble [13] MD is run for 100 ps at T = 300K, followed by 100ps Berendsen NPT ensemble [14] (P = 1bar, T= 300K) MD to help samples reach their initial equilibrium configurations. The samples are then gradually cooled down to 10K at a rate of 1K/ps. Each system is further equilibrated for 100 ps at chosen temperature data points (at P = 1bar) to reduce possible fluctuations.

The temperature-depended densities of both the bulk PE and PE-CNT composites, averaged from simulations of 6 sample sets, are shown in Figure 2. The location of the discontinuity in the slope of the density vs. temperature plot implies the on-set of the glass transition temperature ($T_g$). Experiments have reported $T_g$ for bulk PE varying from 180K to 250K, depending on molecular weights and degrees of cross-linking. [15] In our samples, the computed $T_g$ of about 150K is in agreement with previous MD simulations conducted on similar sized samples. [10] The nanotube-polymer composite, as shown in Fig. 2, tends to have a higher $T_g$ of 170K, as the CNT in the matrix tends to slow the motions of the surrounding molecules below $T_g$. It is possible that the cross-linking of polymer matrix with embedded CNTs might also further reduce the motions of polymer molecules and increase the $T_g$. Recent experiments on epoxy composite with 1% weight CNT indicate an increase of $T_g$ about $10^o C$. [17]

The decrease in the density of the bulk polymer and the composite with increased temperatures indicates thermal expansions of the materials. The slope of the curve at a given temperature gives the volume thermal expansion coefficient at that temperature as $\frac{1}{V}\frac{\partial V}{\partial T}$. It can be noticed that above $T_g$ (170K), the volume thermal expansion increases more rapidly in the composite compared to that in bulk PE. The volume thermal expansion coefficient of the composite is found to be $4.5 \times 10^{-4} K^{-1}$ below $T_g$, about 18% larger than the $3.8 \times 10^{-4} K^{-1}$ of the polymer bulk in the same temperature range. Above $T_g$, the volume thermal expansion coefficient of the composite is found to be $12 \times 10^{-4} K^{-1}$, and is increased as much as 40% compared to $8.6 \times 10^{-4} K^{-1}$ of the bulk PE within the same temperature range.

Our simulations on a larger system have verified this increase of thermal expansion coefficient of the CNT composites. For bulk PE, the sample is consisted of 50 polymer chains, with repeating units $N_p = 100$, in each unit cell (total 5000 atoms). For CNT composites, a 200Å long (10,0) CNT (total 1804 C atoms) is embedded in the PE matrix for each unit cell. The samples are prepared with the similar procedures described above. Figure 3 shows a similar increase of the thermal expansion in the composite at high temperatures. The volume thermal expansion coefficient of the composite is found to be $16.1 \times 10^{-4} K^{-1}$ (T > 400K), $6.9 \times 10^{-4} K^{-1}$ (300K < T < 400K), and $2.2 \times 10^{-4} K^{-1}$ (T < 300K) respectively. For PE matrix, the expansion coefficient is $6.6 \times 10^{-4} K^{-1}$ (T > 300K) and $1.69 \times 10^{-4} K^{-1}$ (T < 300K) respectively. An increase of 142% of the expansion coefficient of the composite can be found, compared to that of the bulk PE at T > 400K. Experiments [18] find the volume expansion coefficient of PE is around $1.0 \times 10^{-4} \text{K}^{-1}$ at room temperature. Our MD simulations on the smaller sized samples show a larger expansion coefficient while the coefficient of the larger sample is more in agreement with experimental data (which is on heavy polymerized systems). For the large sized bulk PE, $T_g$ is found to be around 300K, higher than $T_g$ for small samples discussed above. This is due to the dependence of $T_g$ on molecular weights of polymer chains. Previous MD simulations on long chained PE systems show $T_g$ to be in the range of 230K to 350K, [9] [18] which is similar to our computed $T_g$ for the long chained systems.

The origin of higher thermal expansion coefficients in the composites is explained as follows. The CNT embedded within a polymer matrix has a fixed volume and excludes the occupancy of polymer chains. The enhanced thermal expansions of the composites are attributed to the increase in the excluded volume of embedded CNTs as a function of temperature. To verify this, the MD simulations of the smaller-sized composite sample at 300K are repeated under two constraints: (1) the phonon vibrations within a nanotube are frozen by increasing the stiffness of C-C bonds and by allowing only the motions of the CNT as a whole; (2) in addition to the first, the nanotube as a whole is frozen as well. Results under the second constraint, for a completely frozen CNT, show that the density of the composite reverts back. The density of the composite under the first constraint is intermediate between the completely frozen case and the completely free case. The energy from the VDW interactions is found increased with more allowed freedoms of motions of the CNT (shown in Figure 4). These results suggest that both phonon modes and Brownian motions of the CNT contribute to the excluded volume.

Lastly, in Figure 5, we show the diffusion coefficients $\frac{\partial \langle \Delta r^2 \rangle}{\partial t}$ of the C atoms of the polymer matrix for the small-sized systems of the bulk PE and the composite cases. It is clear that in the composite case, the polymer chains are more diffusive at temperatures above $T_g$. Further more, the increase in the diffusion coefficient parallel to the CNT axis is about 30% larger than the increase perpendicular to the tube axis. This correlates well with the higher thermal expansion coefficient (above $T_g$) data described above, and shows that the increased phonon vibrations and Brownian motions of the CNT are not only coupled efficiently to the motions of the polymer chains but also the coupling is anisotropic. This means that polymer chains parallel to the tube axis will flow or diffuse better at higher temperatures. Recent experiments on CNT-ABS and CNT-RTV composites, by Rick Berrera's group at Rice University, have shown large increases of diffusion coefficients of the matrixes. [19]

The importance of these findings has implications in the processing and applications of CNT reinforced polymer composite materials. The higher diffusion coefficient of the matrix atoms above $T_g$ will allow increased mobility of composite materials during processing steps such as continuous spinning, weaving and extrusion of the materials for fabrication purposes, and also possibly the smoother flow through nozzles for painting or coating applications. The increased thermal expansion and diffusion coefficients, above $T_g$, may make it possible to significantly increase the thermal conductivity of CNT composites at high temperatures.

In summary, we have used MD simulations to study temperature dependent structural behavior of PE-CNT composites. The results show that the thermal expansion coefficient of the composite and diffusivity of polymer molecules increases significantly above glass transition temperature of the composite.

**Acknowledgment.** This work is supported by NASA contract NCC2-5400. DS is supported by NASA contract 704-40-32 to CSC at Ames Research Center.

**Figures**

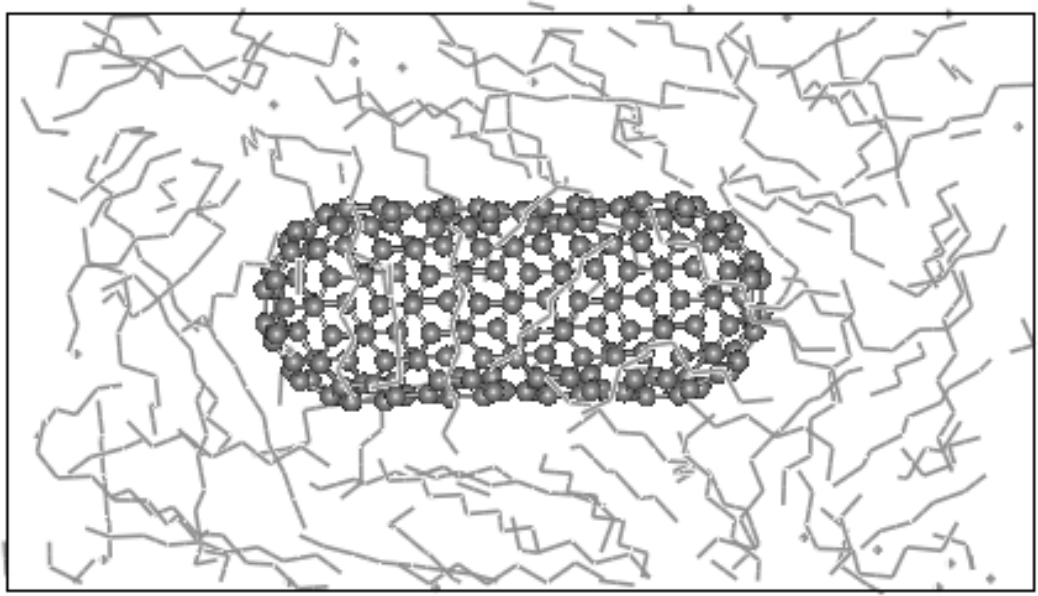

**Figure 1.** The MD simulation unit cell for a composite system with 20 Å long capped (10,0) CNTs embedded in the polyethylene matrix. Periodic boundary condition is used.

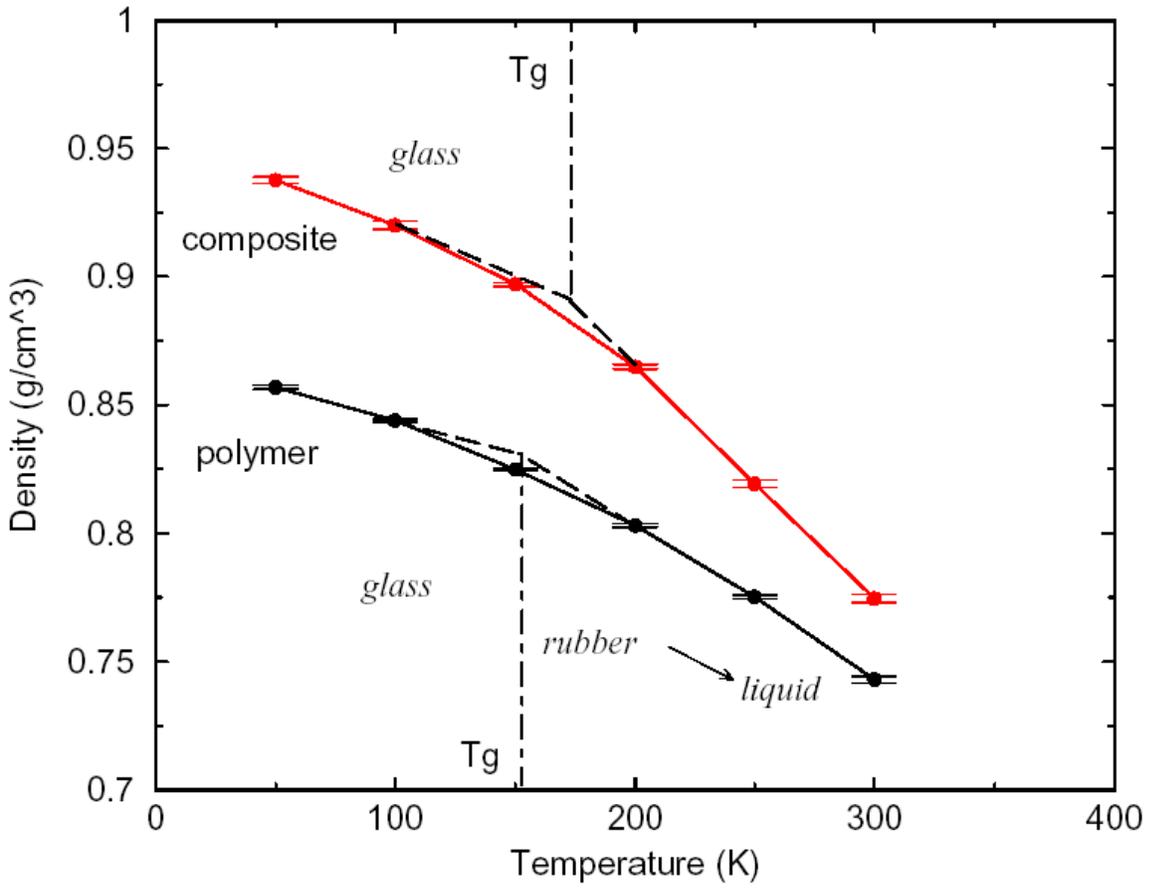

**Figure 2.** The density as a function of temperature for short-chained PE ($N_p$=10) and its composite with 20Å long (10,0) CNTs embedded (averaged from 6 sample sets, the error bars are shown in the plot). Below $T_g$, the systems are in glassy state, and above $T_g$, the systems will be in rubber-like or liquid state, if temperature is higher than melting temperature $T_m$.

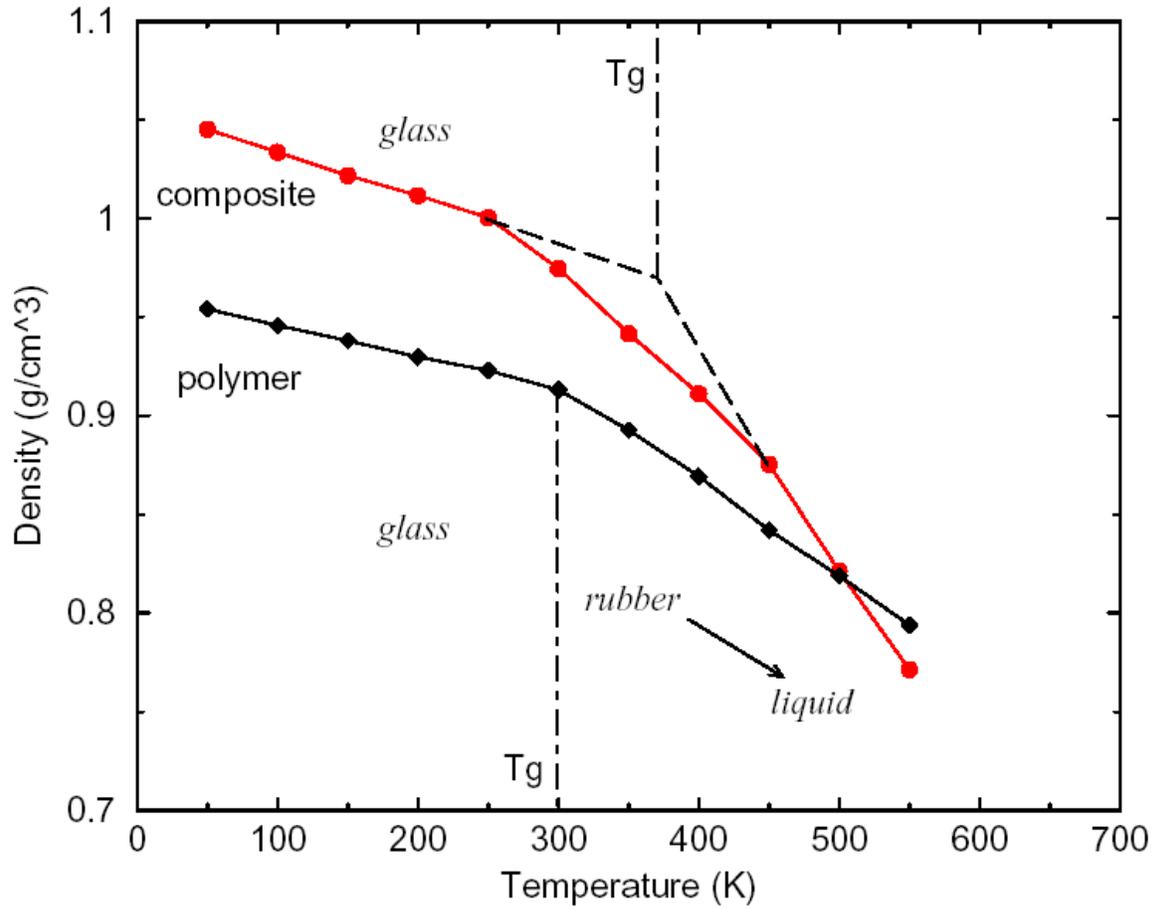

**Figure 3.** The density as a function of temperature for long-chained ($N_p$ = 100) PE and its composite with 200Å long (10,0) CNTs embedded. Below $T_g$, the systems are in glassy state, and above $T_g$, the systems will be in rubber-like or liquid state, if temperature is higher than melting temperature $T_m$.

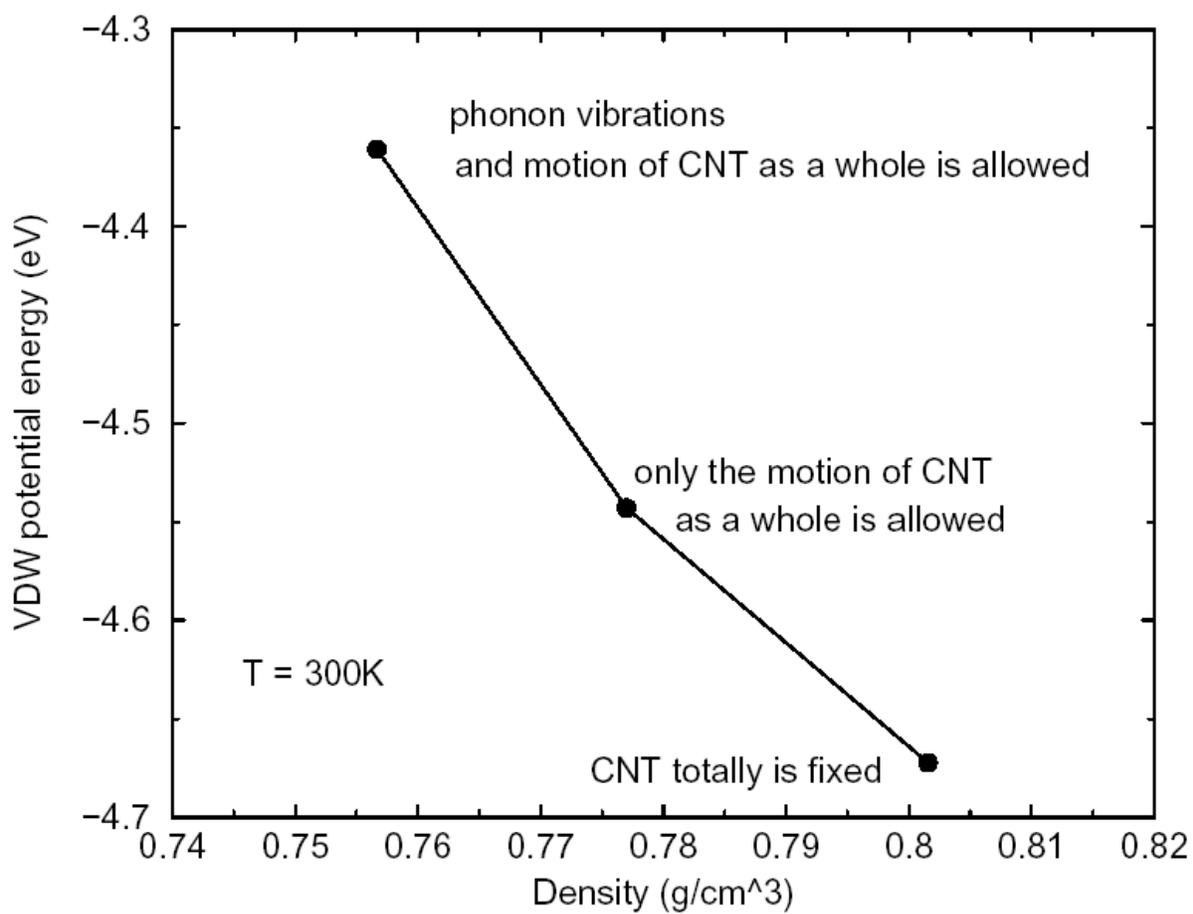

**Figure 4.** The change on density of the CNT composite with phonon modes and Brownian motions frozen for the smaller-sized system at T=300K. Both thermal motions contribute to the exclude volume of embedded CNTs, which lead to the increased thermal expansion. The VDW energy of the system is increased with more allowed freedoms of motions of the embedded CNT.

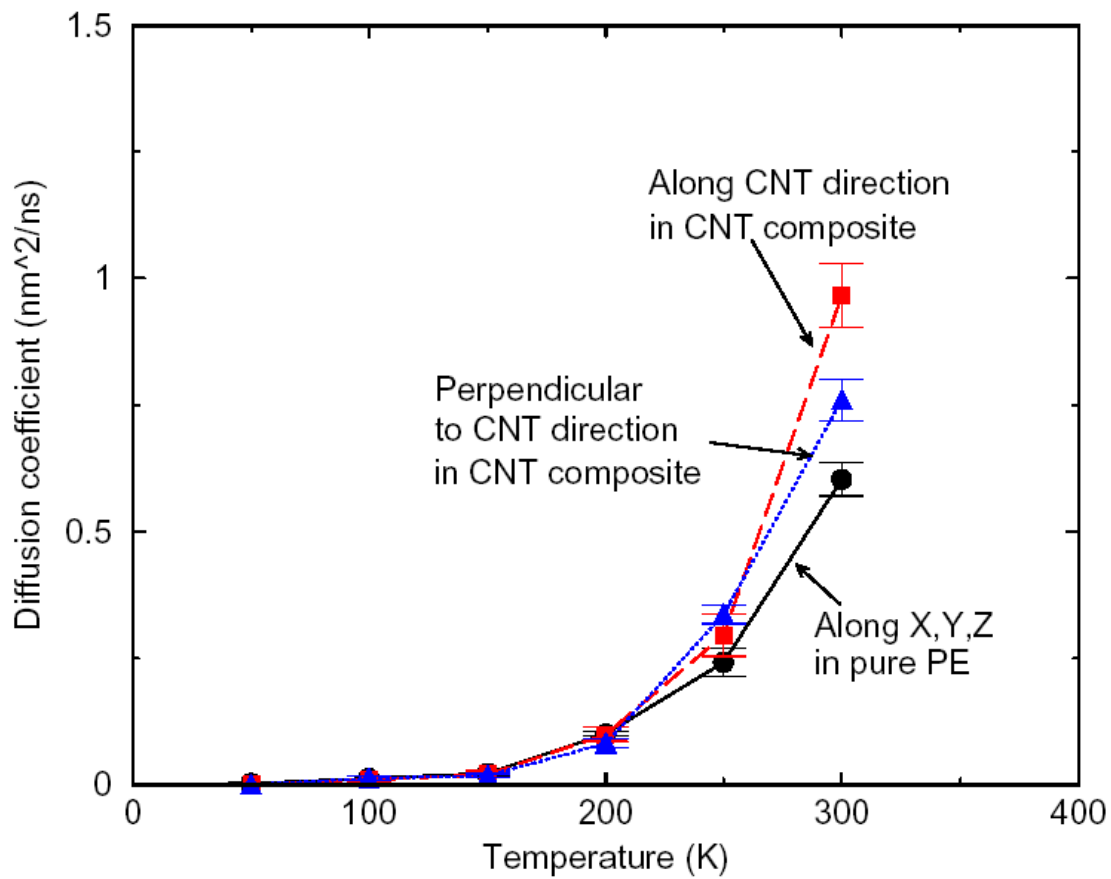

**Figure 5.** The diffusion coefficient of polymer chains in pure PE matrix (short chained systems), and the CNT composite systems. With the presence of CNTs, the diffusion coefficient is increased at high temperatures (T > Tg) and the component parallel to CNT axis has a larger contribution compared with the perpendicular one.